\pdfoutput=1
\documentclass[runningheads]{llncs}
\usepackage[margin=3cm]{geometry}
\usepackage{graphicx}
\usepackage{floatrow}
\usepackage{amsmath}
\usepackage[utf8]{inputenc} 
\usepackage{subcaption}
\usepackage{todonotes}
\usepackage{comment}
\usepackage{booktabs}
\usepackage{pgfplots}

\usepackage{booktabs} 
\usepackage{tikz}
\usetikzlibrary{shapes,arrows}
\usetikzlibrary{positioning}

\usepackage{amsmath,amssymb,graphicx,enumerate,bbm,calc,ifthen,theorem,capt-of}
\newcommand{\assign}{:=}

\newcommand{\mathlambda}{\lambda}
\newcommand{\nobracket}{}
\newcommand{\nocomma}{}

\newcommand{\tmem}[1]{{\em #1\/}}
\newcommand{\tmmathbf}[1]{\ensuremath{\boldsymbol{#1}}}
\newcommand{\tmop}[1]{\ensuremath{\operatorname{#1}}}

\newcommand{\tmfloatcontents}{}
\newlength{\tmfloatwidth}
\newcommand{\tmfloat}[5]{
  \renewcommand{\tmfloatcontents}{#4}
  \setlength{\tmfloatwidth}{\widthof{\tmfloatcontents}+1in}
  \ifthenelse{\equal{#2}{small}}
    {\ifthenelse{\lengthtest{\tmfloatwidth > \linewidth}}
      {\setlength{\tmfloatwidth}{\linewidth}}{}}
    {\setlength{\tmfloatwidth}{\linewidth}}
  \begin{minipage}[#1]{\tmfloatwidth}
    \begin{center}
      \tmfloatcontents
      \captionof{#3}{#5}
    \end{center}
  \end{minipage}}

\begin{document}
\title{A High-Performance Implementation of a Robust Preconditioner for Heterogeneous Problems}

\author{Linus Seelinger\inst{1}
\and
Anne Reinarz\inst{2}
\and
Robert Scheichl\inst{3}
}     

 \authorrunning{L. Seelinger et al.}
%
\institute{Institute for Scientific Computing, Heidelberg University, Heidelberg, Germany\\
\email{linus.seelinger@iwr.uni-heidelberg.de}\and
Department of Informatics, Technical University of Munich, Garching, Germany \\
\email{reinarz@in.tum.de} \and
Institute for Applied Mathematics, Heidelberg University, Heidelberg, Germany\\
\email{r.scheichl@uni-heidelberg.de}}
\maketitle              

\begin{abstract}

We present an efficient implementation of the highly robust and scalable \mbox{GenEO}  preconditioner \cite{Spillane_2011} in the high-performance PDE framework DUNE \cite{PDELAB}. The GenEO coarse space is constructed by combining low energy solutions of a local generalised eigenproblem using a partition of unity. In this paper we demonstrate both weak and strong scaling for the GenEO solver on over $15,000$ cores by solving an industrially motivated problem with over 200 million degrees of freedom. Further, we show that for highly complex parameter distributions arising in certain real-world applications, established methods become intractable while GenEO remains fully effective.
The purpose of this paper is two-fold: to demonstrate the robustness and high parallel efficiency of the solver and to document the technical details that are crucial to the efficiency of the code.


\end{abstract}

 \parindent0pt

\section{Introduction}
Computer simulations have become a vital tool in science and engineering. The
demand for solving PDEs on ever larger domains and increasing accuracy necessitates the use of high performance computers and the implementation of efficient parallel algorithms.
When designing parallel algorithms two issues are crucial: robustness and scalability.
\renewcommand{\labelenumi}{\roman{enumi})}
\begin{enumerate}
 \item \textbf{Robustness:} The parameters involved in the PDE affect the performance of the algorithm to a large extent. A frequent issue is a distribution of parameters with a large contrast and that contain jumps at different length scales. In many cases a large contrast leads to very slow convergence or even stops it completely.
 \item \textbf{Scalability:}  The immediate scalability of the finite element method is limited as each degree of freedom is coupled with all others.
\end{enumerate}

One approach to achieve scalability in solving partial differential equations
are domain decomposition methods (see e.g. {\cite{SmithGroop,ToselliWidlund}}), splitting the given domain into 
multiple subdomains. The solution of the original problem restricted to
each subdomain is computed in parallel and the results are combined to form
an approximate solution. This is repeated until convergence is reached. The number of these iterations, however, still depends
strongly on the number of subdomains involved as well as coefficient
variations. Introducing an additional coarse space that covers all subdomains
can restore performance for large numbers of subdomains.

This global space can be tailored to specific problem, as in the generalized finite element method  {\cite{BabuskaCalozOsborn}}. While these methods are applicable to an entire class of parameter distributions, each of these approaches is based on certain assumptions on the  parameters, e.g. the parameters vary strongly only in one direction.
The GenEO coarse space chosen in this work is a related approach, originally
introduced in {\cite{Spillane_2011}}. It does not require a-priori knowledge of the parameter distribution and is applicable to a wide range of problems making it suitable as a 'black-box' solver.


In this paper we focus on two different elliptic problems; the Darcy equation describing incompressible flow in a porous medium and the anisotropic linear elasticity equations. For both equations the case of heterogeneous coefficients is of great interest.

Composite materials, which make up over 50\% of recent aircraft constructions, are manufactured from carbon fibres and soft resin layers. The large jump in material properties between the layers makes the simulation of these materials challenging. Commercial solvers such as ABAQUS often rely on direct solvers to deal with these jumps \cite{dunecomp}. However, the scalability of direct solvers is limited. We will demonstrate that the GenEO approach converges independentally of the contrast in material properties and the number of subdomains.


The paper is structured as follows. We will sketch the construction of the GenEO preconditioner. Then we will discuss how to efficiently implement the solver in the high-performance finite element framework DUNE \cite{DUNE1,DUNE2}. Finally we will provide several numerical experiments demonstrating both the robustness and scalability of the solver, including one large-scale industrially motivated example.

\section{Problem formulation and variational setting}
Let $V$ be a Hilbert space, $a\,:\,V\times V \rightarrow \mathbb{R}$  a symmetric and coercive bilinear form
and $f\in V'$. We consider the following abstract variational problem.
Find $v\in V$ such that
\begin{equation}\label{eq-abstractsetting}
 a(v,w) =\langle f, w \rangle,~~~~\forall w \in V,
\end{equation}
where $\langle\cdot,\cdot\rangle$ denotes the duality pairing.

 This variational problem is associated with an elliptic boundary value problem on a domain $\Omega \subset \mathbb{R}^d,$
 $d= 2,3$ with Dirichlet boundary $\partial \Omega_D$. In particular, we focus on the following two examples.

 \begin{enumerate}
  \item  {\bf Darcy problem: } Given material properties $\kappa \in L^\infty(\Omega)$,
  find $v\in V = \{v\in H^1(\Omega)\,:\, v|_{\Omega_D} = 0\}$ such that
 \begin{equation}\label{eq-darcy}
   a(v,w) = \int_{\Omega} \kappa ( x) \nabla v ( x) \cdot \nabla w ( x) \, d x  = \int_{\Omega} f( x) w ( x) \, d x ,
   ~~~~~ \forall w \in V.
 \end{equation}

\item { \bf Linear Elasticity:  } Given material properties $C$, find $v\in V = \{ v \in H^1(\Omega)^d \, : \, v|_{\Omega_D} = 0\}$ such that
 \begin{equation}\label{eq-linearelasticity}
   a(v,w) = \int_{\Omega} C( x)\varepsilon( v) : \varepsilon(w) d x = \int_\Omega  f \cdot w \, d x
+ \int_{\partial \Omega} (\sigma \cdot { n}) \cdot{ v}\,d x,~~~~~ \forall w \in V,
 \end{equation}
where $\varepsilon_{ij}(v)=\frac{1}{2}(\partial_i v_j + \partial_j v_i)$ is the strain, and
$\sigma_{ij}(v)=\sum_{k,l=1}^dC_{ijkl}\varepsilon_{kl}$ is the stress.
 \end{enumerate}


Consider a discretization of the variational problem (\ref{eq-abstractsetting}) using finite elements on a mesh $T_h$ of $\Omega$  such that $\overline{\Omega} = \cup_{\tau\in T_h}\tau.$ Let $V_h \subset V$ be a conforming space of finite element functions. 
Then the discrete form of (\ref{eq-abstractsetting}) is: Find $v_h\in V_h$ such that
\begin{equation}
  a(v_h,w_h) =\langle f, w_h \rangle,~~~~\forall w_h \in V_h.
\end{equation}

\section{The GenEO Preconditioner}\label{geneo-prec}
In order to leverage the potential of modern high performance computers,
parallelization is crucial. The task must be split into pieces that
can be computed independently, and communication between processes must be
minimized. However, in the finite element method  each degree of freedom is indirectly coupled with all others. In  the overlapping additive Schwarz method parallelization is achieved by splitting the computational domain $\Omega$ into multiple overlapping overlaps subdomains and solving a local problem on each subdomain. In an iterative procedure results from each
subdomain are added on the overlaps using nearest-neighbor communication and an updated local problem is solved taking into account new information from  neighboring subdomains.

We generate the overlapping subdomains by starting from a non-overlapping
subdivision $\{ \Omega_j^{'} \}_{j = 1}^N$ of $\Omega$. Each $\Omega_j^{'}$ is the union
of mesh elements from $\mathcal{T}_h$. An arbitrary number of layers of elements can be added to each $\Omega_j^{'}$ by applying the definition multiple times, resulting in
overlapping subdomains $\Omega_j$.
  For each subdomain $1 \leqslant j \leqslant N$, the overlapping zone is
  defined as
  \[ \Omega_j^o \assign \{ x \in \Omega_j : \exists j' \neq j \tmop{such}
     \tmop{that} x \in \Omega_{j'} \} . \]
  We denote the restriction of the function space  $V_h$ to $\Omega_j$ by
  $ V_h ( \Omega_j) \assign \{ v | \nobracket_{\Omega_j} : v \in V_h \}$ for each $1 \leqslant j \leqslant N$ and the restiction of $V_h(\Omega_j)$ to  functions that inherently fulfill a homogeneous Dirichlet condition on their respective subdomain by $V_{h,0}(\Omega_j)$.

\begin{definition}
  We define the prolongation operator
  $R_j^T : V_{h, 0} ( \Omega_j) \rightarrow V_h $
  for each element $v_j \in V_{h, 0} ( \Omega_j)$ as
  $ R_j^T v_j | \nobracket_{\Omega_j} = v_j $
  and
 $R_j^T v_j | \nobracket_{\Omega \backslash \Omega_j} = 0.$
 The corresponding restriction operator by $R_j$ is defined as
  \begin{eqnarray*}
    \langle R_j g, v \rangle = \langle g, R_j^T v \rangle, & ~~ &
    \forall v \in V_{h, 0} ( \Omega_j) \nocomma, g \in V_h' .
  \end{eqnarray*}
\end{definition}


We denote the matrix form of the restriction operators $R_j$ by
$\tmmathbf{R}_j$ and of the system matrix by $\tmmathbf{A}$. Further, we denote the problems restricted to the subdomains by $\tmmathbf{A}_j  \assign \tmmathbf{R}_j \tmmathbf{A}\tmmathbf{R}_j^T$, for all $j = 1, \ldots, N$.  Then the additive Schwarz preconditioner is given by
  \[ \tmmathbf{M}_{\tmop{AS}, 1}^{- 1} \assign \sum_{j = 1}^N \tmmathbf{R}_j^T
     \tmmathbf{A}_j^{- 1} \tmmathbf{R}_j . \]

Due to local exchange of information, the number of iterations required tends
to increase strongly with the number of subdomains involved.
This can be overcome by additionally solving a suitable
global coarse problem. The resulting method is referred to as a two-level additive Schwarz method.  For the coarse space $V_H$, denote the natural embedding by $R_H^T : V_H
  \rightarrow V_h$ and its adjoint by $R_H$.

  \begin{definition}[Two-level Additive Schwarz]
Denote the problem restricted to the coarse space by $\mathbf{A}_H\assign \tmmathbf{R}_H \tmmathbf{A}\tmmathbf{R}_H^T,$ for all $j = 1, \ldots, N$. Then the two-level preconditioner is given by:
  \[ \tmmathbf{M}_{\tmop{AS}, 2}^{- 1} \assign \tmmathbf{R}_H^T
     \tmmathbf{A}_H^{- 1} \tmmathbf{R}_H + \sum_{j = 1}^N \tmmathbf{R}_j^T
     \tmmathbf{A}_j^{- 1} \tmmathbf{R}_j . \]
\end{definition}

The analysis framework from \cite{ToselliWidlund} allows for upper and lower bounds on the condition number of the preconditioned system.  The largest eigenvalue of the preconditioned system can be bounded using the maximum number of subdomains that cover each point $k_0$. Clearly $k_0$ can easily be controlled by constructing a
reasonable domain decomposition. The bound on the smallest eigenvalue depends on the stable splitting constant. Thus, this constant should ideally be small. The GenEO coarse space was designed to minmise the stable-splitting constant, it was first introduced in {\cite{Spillane_2011}}, followed by a full theoretical analysis in {\cite{Spillane_2014}}. 

A key ingredient in the GenEO coarse space is the partition of unity, which
allows 'stitching together' the local basis results on each subdomain to form
a suitable basis of the entire domain. 
\begin{definition}[Partition of unity]
  Given weights $\mu_{j, k} \in [0,1]$ with $ \sum_{1 \leq j \leq N}  \mu_{j, k} = 1$, the
  associated partition of unity operator for each
  subdomain $1 \leqslant j \leqslant N$ is defined by
  \[ \Xi_j ( v) \assign \sum_{k \in \tmop{dof} ( \Omega_j)} \mu_{j,
     k} v_k \phi_k | \nobracket_{\Omega_j},
  \text{ for any } v \in V_h ( \Omega_j).\]
  Here we denote the value of $v$ at degree of freedom $k$ as $v_k$, and by $\phi_k$ the basis function associated to the same degree of freedom.
\end{definition}

\begin{figure}[t]
 \centering
 \caption{Plot of two different partitions of unity. Left: the standard piecewise constant partition of unity. Right: Sarkis partition of unity.}
  \includegraphics[width=.2\linewidth]{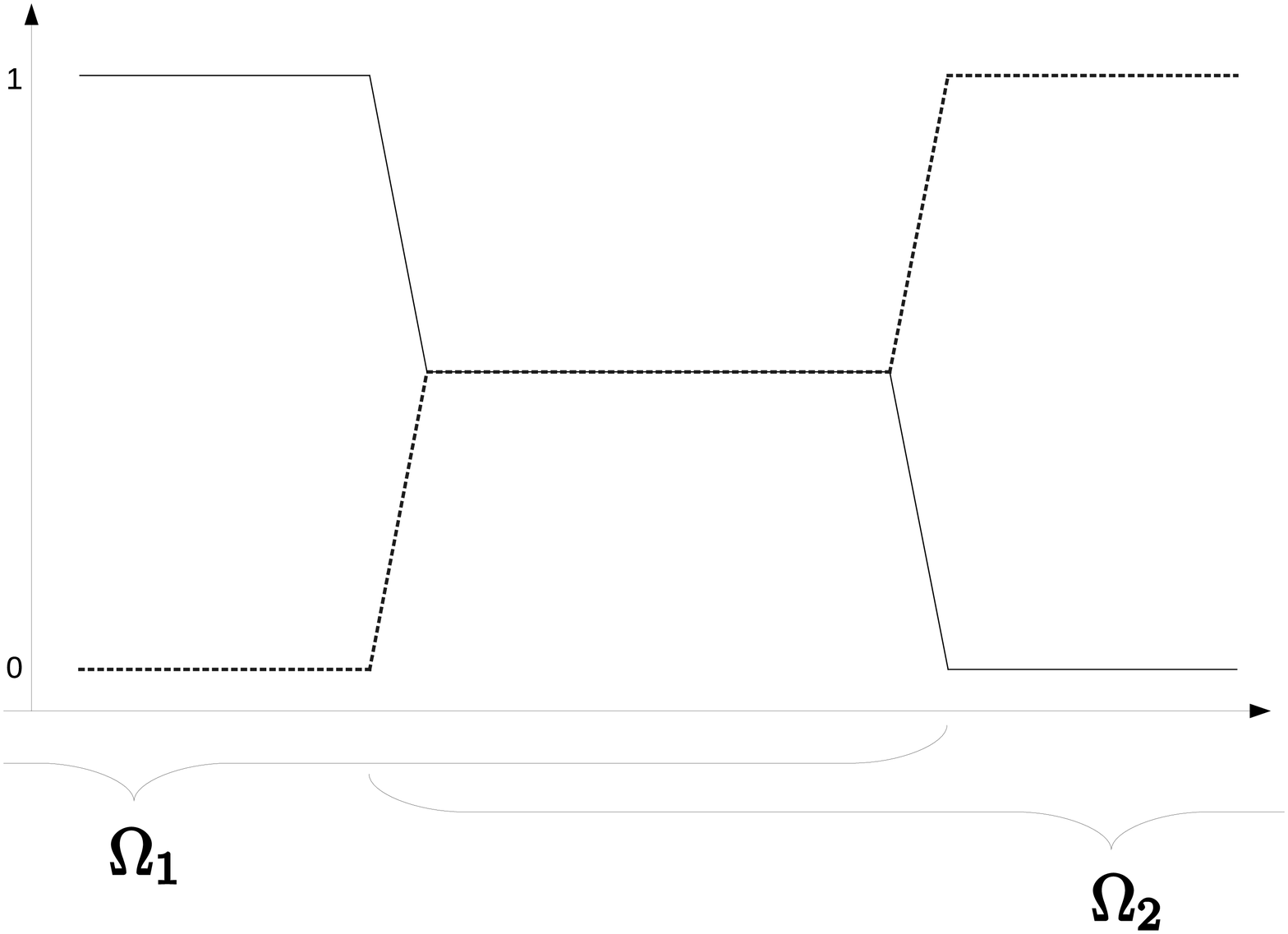}
  \includegraphics[width=.28\linewidth]{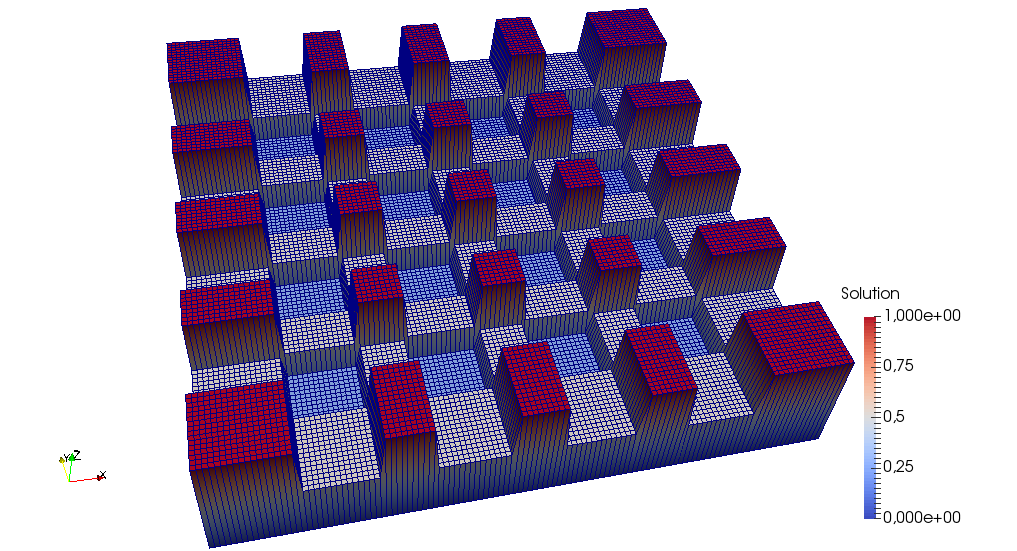}
  \includegraphics[width=.2\linewidth]{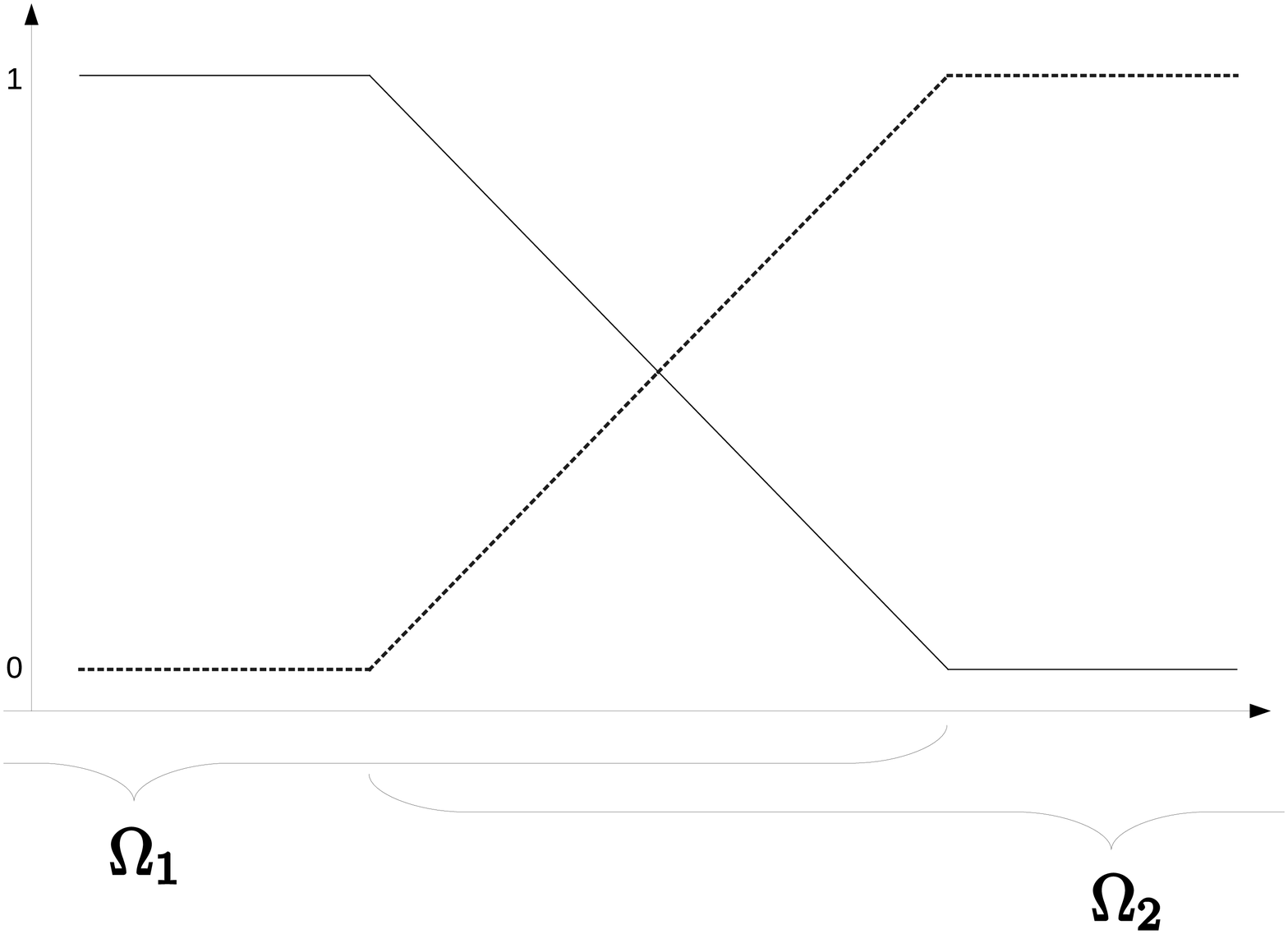}
  \includegraphics[width=.28\linewidth]{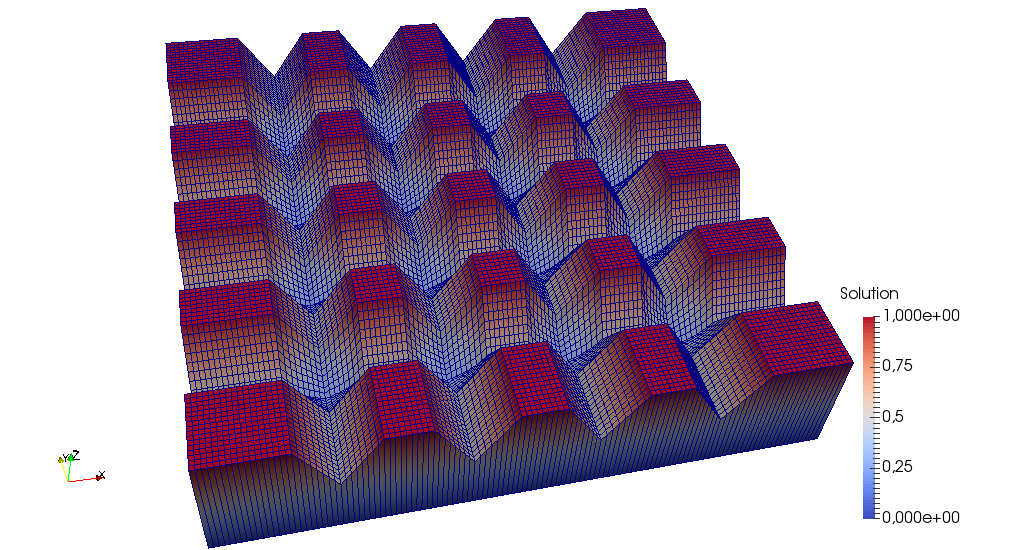}
\label{fig:partunity}
\end{figure}

Figure \ref{fig:partunity} shows two examples of partitions of unity. 
It should be noted that implementing a smooth partition of unity for arbitrary subdomain arrangements can be challenging and in our tests we have not found this to have a large influence on the efficiency of the preconditioner.
 Using the partition of unity operators defined above, we can now state the
 definition of the GenEO coarse space. The space is
 constructed from the eigenvectors of a specific generalized eigenproblem
 representing the inequality required by the stable splitting.

  For each subdomain $j = 1, \ldots, N$, we define the generalized
  eigenproblem: Find $p\in V_h(\Omega_j)$ such that
  \begin{equation}\label{GenEOEigproblem}
    a_{\Omega_j} ( p, v) = \lambda a_{\Omega_j^o} ( \Xi_j ( p), \Xi_j ( v)),~~ \forall v \in V_h ( \Omega_j).
  \end{equation}
Note that the eigenproblems are local to their respective subdomain
$\Omega_j$, i.e. they can be computed in parallel. To use them as a global basis they need to be extended to the entire domain using the partition of unity operators.

\begin{definition}[GenEO coarse space]
  \label{GenEOSpace}
  For each subdomain $j = 1, \ldots, N$, let $( p_k^j)^{m_j}_{k = 1}$ be the
  eigenfunctions from the eigenproblem in \eqref{GenEOEigproblem}
  corresponding to the $m_j$ smallest eigenvalues. Then the GenEO coarse space
  is defined as
  \[ V_H \assign \tmop{span} \{ R_j^T \Xi_j ( p_k^j) : k = 1, \ldots, m_j ; j
     = 1, \ldots, N \} . \]
\end{definition}



In \cite{Spillane_2014} the following bound on the condition number of the matrix has been shown.
\begin{theorem}
  \label{ConditionNumberBound}

  For all $1 \leqslant j \leqslant N$, let the number of eigenvectors chosen in each subdomain be
  \[ m_j \assign \min \left\{ m : \lambda_{m + 1}^j > \frac{\delta_j}{H_j}
     \right\}, \]
  where $\delta_j$ is a measure of the width of the overlap $\Omega_j^o$ and
  $H_j = \tmop{diam} ( \Omega_j)$. Then,
  \[ \kappa ( \tmmathbf{M}^{- 1}_{\tmop{AS}, 2} \tmmathbf{A}) \leqslant ( 1 + k_0) \left[ 2 + k_0 (
     2 k_0 + 1) \max_{1 \leqslant j \leqslant N} \left( 1 +
     \frac{H_j}{\delta_j} \right) \right] . \]
\end{theorem}
Thus the algorithm presented is provably independent of the number of subdomains and coefficient variations in the problem. Note that the choice of threshold by which the eigenvectors are selected is not unique. The threshold can be scaled to vary the condition number of the preconditioned system while retaining robustness. This allows us to control the total number of preconditioned CG iterations.

\section{HPC Implementation of GenEO in Modern PDE Frameworks}

When implementing the GenEO preconditioner in a PDE framework, the primary
goal is to preserve the beneficial properties offered by its theoretical construction, namely:
\renewcommand{\labelenumi}{\roman{enumi})}
\begin{enumerate}
\item \textbf{High parallel scalability: } Since  the condition bound in Thm. \ref{ConditionNumberBound} is independent of the number of subdomains we expect the implementation to yield high parallel scalability. The  solution of the eigenproblems  parallelizes trivially. However, care has to be taken when it comes to the communication necessary to set up the coarse matrix.

\item \textbf{Robustness with respect to problem parameters: }
While this is an inherent property of the preconditioner, some care is required in implementing the Dirichlet boundary conditions.

\item \textbf{Applicability to various types of PDEs: }
The theoretical framework only requires a symmetric positive definite bilinear form as in (\ref{eq-abstractsetting}). This flexibility can be preserved in any numerical framework that is based on abstract bilinear forms. This is the case for many modern PDE frameworks, e.g. FEniCS \cite{fenics}, DUNE \cite{DUNE1}, or deal.ii \cite{deal}.
\end{enumerate}

In this section, we present a new implementation of the GenEO coarse space and preconditioner within DUNE (Distributed and Unified Numerics Environment), which fulfills these properties. This serves as a reference for the implementation, which is freely available as part of the \texttt{dune-pdelab} module \cite{PDELAB} since version 2.6, as well as a general guideline for future implementations in other software packages. DUNE is a generic package that provides the user with key ingredients for solving any FEM problem. As an open source framework written using modern C++ programming techniques, it allows for modularity and reusability while providing HPC grade performance.

\subsection{Prerequisites}
Many of the components required to implement a two-level Schwarz method already exist within DUNE. In particular, we use the PDELab discretization module's functionality to assemble stiffness matrices based on bilinear forms and for efficient communication
across overlapping subdomains. The GenEO basis functions have support not restricted to individual elements, which makes the existing high-level components
of PDELab unsuited for storing the coarse space. As part of this project, components facilitating such coarse spaces were fully integrated within the framework. Further, an efficient sequential solver for generalized eigenproblems is needed. Here, we choose ARPACK \cite{Lehoucq97arpackusers}.

\subsection{General Structure}
\begin{figure}[t]
\caption{Class hierarchy of GenEO implementation in DUNE PDELab}
\begin{subfigure}{.49\textwidth}
\tikzstyle{block} = [rectangle, draw,
    text width=17em, text centered, rounded corners, minimum height=3em]
\tikzstyle{line} = [draw, -latex']
\centering
\begin{tikzpicture}[node distance = .6cm, auto]
    \node [block] (2lvlSchwarz) {TwoLevelOverlappingAdditiveSchwarz};
    \node [block, above= of 2lvlSchwarz] (CoarseSpace) {CoarseSpace};
    \node [block, above= of CoarseSpace] (SubdomainBasis) {SubdomainBasis};
    \node [block, above= of SubdomainBasis] (PartitionOfUnity) {PartitionOfUnity};

    \path [line] (2lvlSchwarz) -- (CoarseSpace);
    \path [line] (CoarseSpace) -- (SubdomainBasis);
    \path [line] (SubdomainBasis) -- (PartitionOfUnity);
\end{tikzpicture}
\caption{Abstract hierarchy}
\end{subfigure}
\begin{subfigure}{.49\textwidth}
\centering
\tikzstyle{block} = [rectangle, draw,
    text width=17em, text centered, rounded corners, minimum height=3em]
\tikzstyle{line} = [draw, -latex']
\begin{tikzpicture}[node distance = .6cm, auto]
    \node [block] (2lvlSchwarz) {TwoLevelOverlappingAdditiveSchwarz};
    \node [block, above= of 2lvlSchwarz] (SubdomainProjectedCoarseSpace) {SubdomainProjectedCoarseSpace};
    \node [block, above= of SubdomainProjectedCoarseSpace] (GenEOBasis) {GenEOBasis};
    \node [block, above= of GenEOBasis] (StandardPartitionOfUnity) {StandardPartitionOfUnity};

    \path [line] (2lvlSchwarz) -- (SubdomainProjectedCoarseSpace);
    \path [line] (SubdomainProjectedCoarseSpace) -- (GenEOBasis);
    \path [line] (GenEOBasis) -- (StandardPartitionOfUnity);
\end{tikzpicture}
\caption{Specific setup for GenEO preconditioner}
\end{subfigure}
\label{class_hierarchy}
\end{figure}
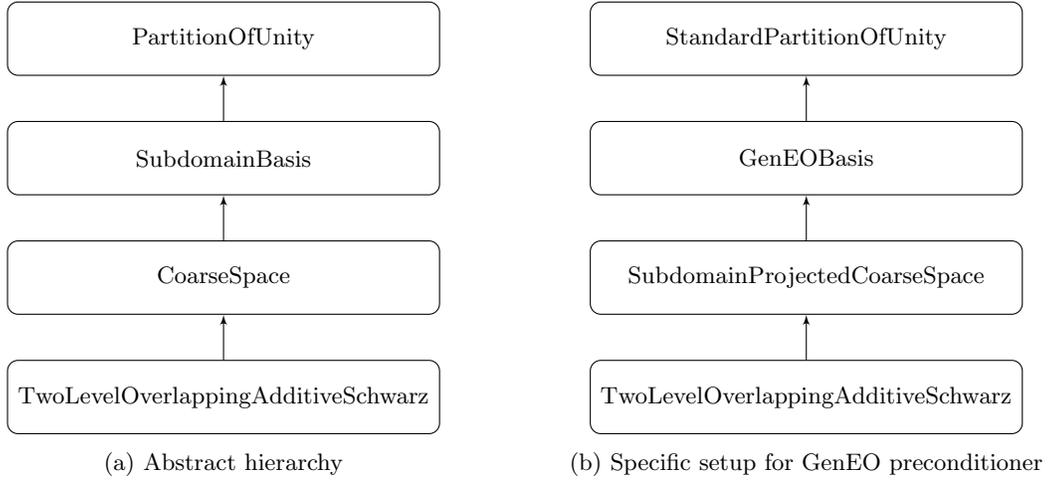

The implementation in PDELab closely follows the structure of the previous section. All mathematical objects are
represented as individual classes (see Fig.~\ref{class_hierarchy}). This separation of concerns leads to an easy to understand and well-structured code.
Further, components are easily interchangeable when constructing related methods.
In particular, the intricate process of constructing a global coarse space from per-subdomain basis functions is entirely contained in the class  \texttt{SubdomainProjectedCoarseSpace}. Thus, the GenEO basis can easily be replaced by a different approach, as only the local basis functions need to be defined on that level.

\subsection{Discrete Basis}
To calculate GenEO basis functions we solve the discrete form of the eigenproblem in Def. \ref{GenEOEigproblem}, i.e.
\[ \tmmathbf{A}_j p_k^j = \lambda_k^j \tmmathbf{X_j}  \tmmathbf{A}_j^o \tmmathbf{X}_j p_k^j, \]
where $\tmmathbf{A}_j^o$ is the discretisation matrix assembled in the overlap region and $\tmmathbf{X}_j$ is the discrete form of the partition of unity.

The matrix $\tmmathbf{A}_j$ has to be assembled with Dirichlet constraints on the
domain boundary as prescribed by the given PDE problem. However, in contrast
to the matrices needed for the one-level component of the two-level additive Schwarz method,
no Dirichlet constraints are imposed on subdomain boundaries.


For assembling $\tmmathbf{A}_j^o$, the same boundary conditions can be
applied. However, additionally, the matrix should only be assembled on the
overlap region. 
Internally these elements are
determined by adding a vector of ones across subdomains and checking for results greater than one.


The matrices $\tmmathbf{X}_j$ representing the partition of unity operator are diagonal
and can be stored as vectors. Entries of
$\tmmathbf{X}_j$ corresponding to Dirichlet domain boundaries or processor boundaries
should be zero, and in sum they should add up to one across subdomains.
Such a partition of unity is generated by adding vectors of ones  with one communication between subdomains. 

\subsection{Solving the Eigenproblem}
As the eigenproblems are defined per-subdomain, the eigensolver itself does not need
to run in parallel. However, solving larger problems requires an efficient iterative
solver. A suitable choice is ARPACK \cite{Lehoucq97arpackusers}.
As the eigenvalues of interest for the GenEO coarse space are those of
smallest magnitude, the {\tmem{Shift and Invert Spectral Transformation Mode}}
supported by ARPACK is used. Instead of the generalized
eigenproblem $ \tmmathbf{A}x = \tmmathbf{M}x \lambda$, ARPACK solves the transformed problem
$ ( A - \sigma M)^{- 1} M x = x \nu$.
The eigenvalues of the transformed problem are related to those of the
original problem by $\nu = \frac{1}{\mathlambda - \sigma}$ and the eigenvectors
are identical. In the transformed problem, the eigenvalues of the original
problem whose absolute values are closest to $\sigma$ are now the eigenvalues
of largest magnitude, and can therefore be efficiently solved by the Krylov
method. Choosing $\sigma$ near zero, the method delivers the eigenvalues of smallest
magnitude at good performance.
Finally, in order to form the actual basis vectors, the eigenvectors are
multiplied by $\tmmathbf{X}_j$ and then normalized in the $l^2$ norm, as ARPACK delivers vectors of
strongly varying norms.


\subsection{Scalable Coarse Setup}

Assembling the coarse matrix $\tmmathbf{R}_H \tmmathbf{A}_H\tmmathbf{R}_H^T$ requires particular care, as it is a non-localized, not trivially scalable operation.
Due to domain decomposition, the global matrix $\tmmathbf{A}$ is only available in distributed form as matrices $\tmmathbf{A}_j$.
Exploiting local support of basis functions, the coarse matrix $\tmmathbf{A}_H$ breaks down into \[ ( \tmmathbf{A}_H)_{i, j} = ( \tmmathbf{R}_H \tmmathbf{A}_H\tmmathbf{R}_H^T)_{i, j} = \varphi_i \tmmathbf{A}_i \varphi_j. \]
We note that $\varphi_i \tmmathbf{A}_i \varphi_j$ is zero for $\Omega_i \cap
\Omega_j = \varnothing$, leading to a sparse structure in $A_H$. Therefore,
all rows $i$ of $\tmmathbf{A}_H$ associated to basis functions $\varphi_i$ can be computed
on the associated process locally while only requiring basis
functions $\varphi_j$ from adjacent subdomains.
In the implementation
multiple basis functions are communicated in a single step.

The resulting blocks are combined into a matrix $\tmmathbf{A}_H$ available on all processes, using direct MPI calls, while exploiting sparsity.
Communication effort obviously increases with the
dimension of $V_H$. This is a direct consequence of how two-level preconditioners are designed, and a good balance between coarse space size and preconditioner performance must be found.

The restriction and prolongation operators $\tmmathbf{R}_H$ and $\tmmathbf{R}_H^T$ are also only available locally. In case of the restriction $\tmmathbf{R}_H v_h$ of a distributed vector $v_h \in V_{h, 0} ( \Omega)$, it holds
\[ ( \tmmathbf{R}_H v_h)_i = \varphi_i \cdot v_h . \]
Each row $i$ can be computed by the process associated to $\varphi_i$, and
the rows can be exchanged among all processes via {\tmem{MPI\_Allgatherv}}.
Again, the communication effort increases with the dimension of $V_H$.

Finally, the prolongation $\tmmathbf{R}_H^T v_H$ of a global vector $v_H \in
V_H$ fulfills \[ \tmmathbf{R}_H^T v_H = \sum_i \varphi_i  ( v_{H_{}})_i . \]
Here, each part of the sum associated with a processor can be computed
locally 
and combined by nearest-neighbor communication, scaling ideally.

\section{Numerical Experiments}
In this section  we demonstrate the solvers salient features, including its high parallel scalability up to  $15,360$ cores, its robustness to heterogeneous material parameterss and its applicability to different elliptic PDEs. With exception of the final large-scale experiment all numerical examples in this section have been computed using the Balena HPC cluster of the University of Bath. Balena consists of $192$ nodes each with two $8$-core Intel Xeon E5-2650v2 Ivybridge processors,
each running at $2.6$ GHz and giving a total of $3072$ available cores.

\subsection{GenEO Basis on Highly Structured Problems}
\label{GenEOStructuredProblems}

\begin{figure}[t]
 \centering
  \includegraphics[width=.24\linewidth]{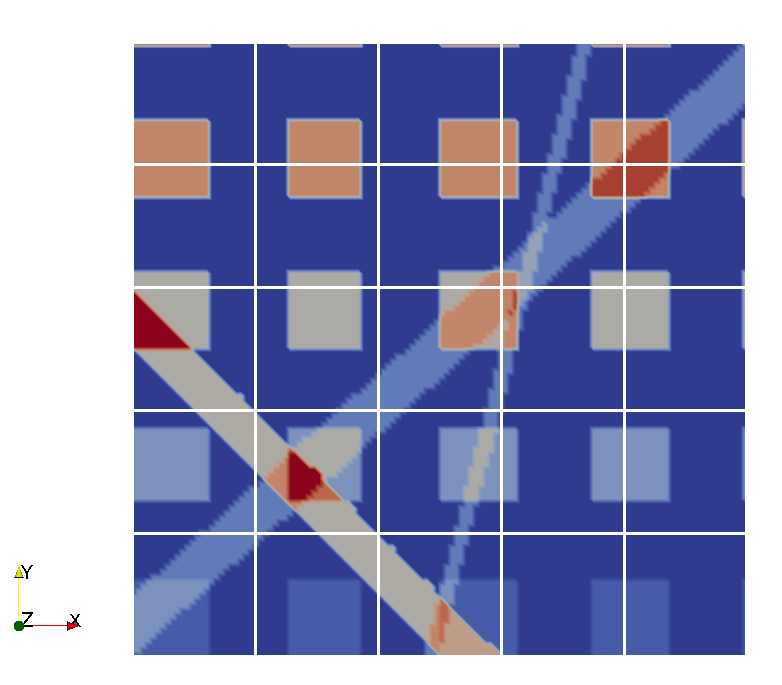}
  \includegraphics[width=.24\linewidth]{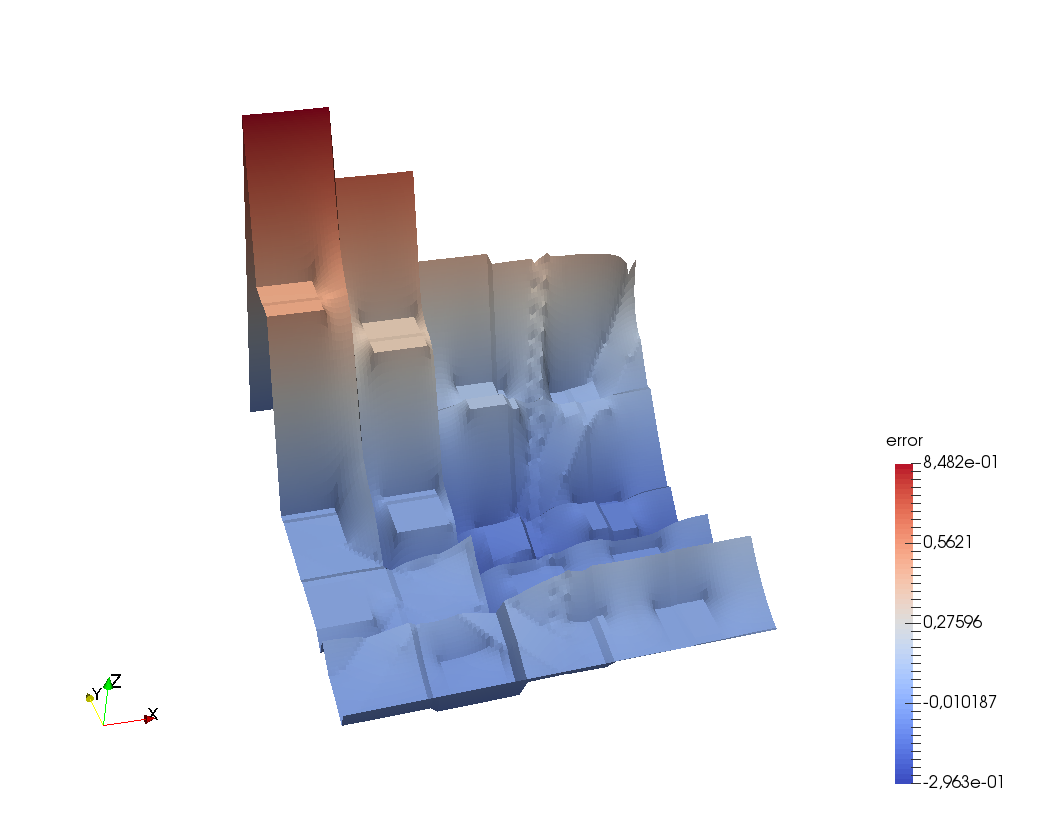}
  \includegraphics[width=.24\linewidth]{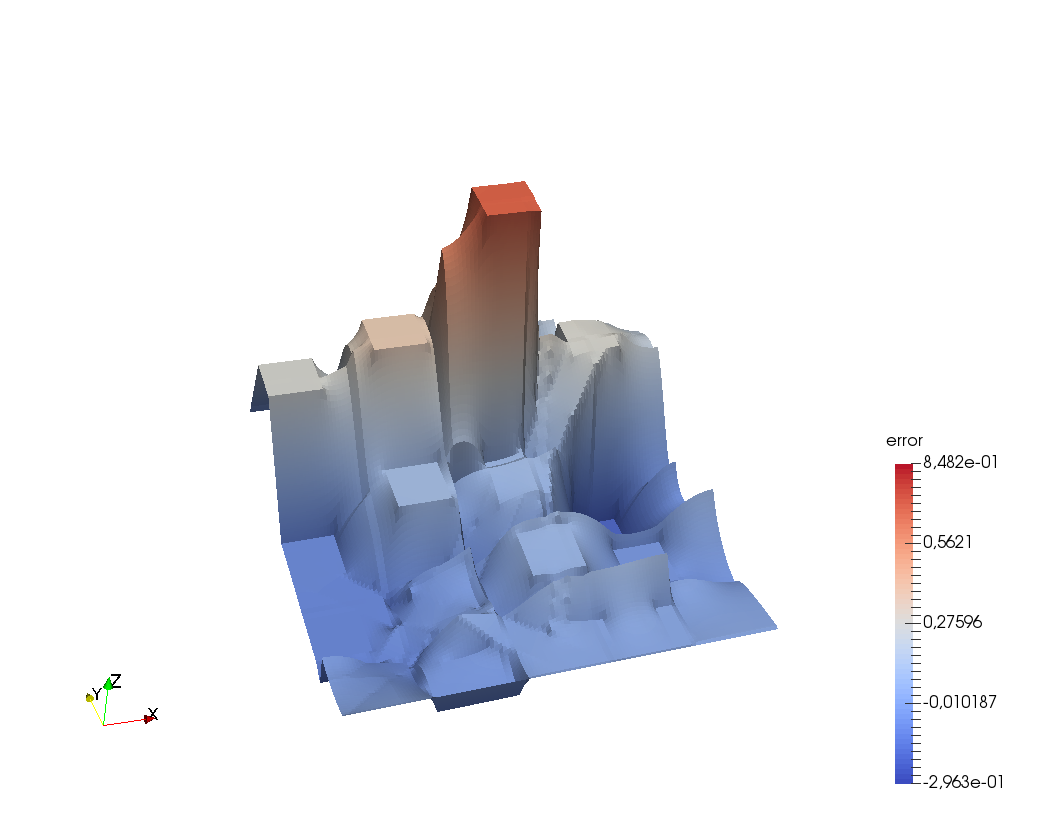}
  \includegraphics[width=.24\linewidth]{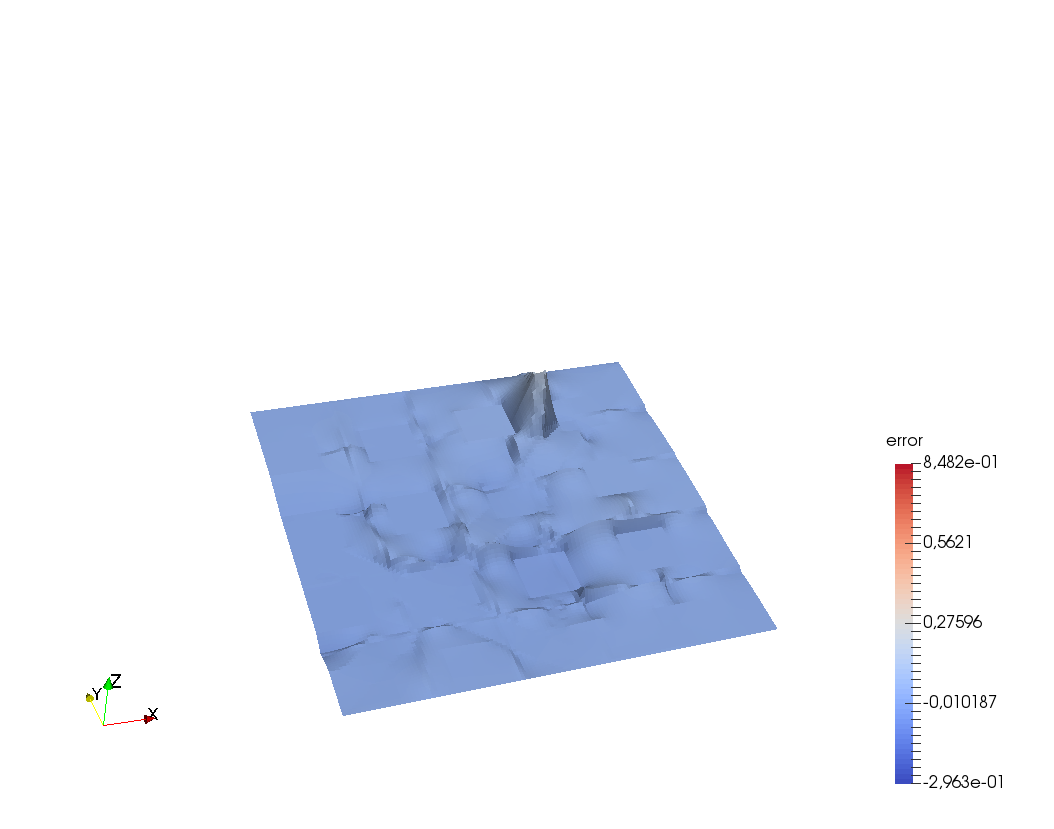}
 \caption{Coarse approximation error. From left to right: the parameter distribution and domain decomposition, the error $u - u_H$ with $1$, $2$ and $4$ eigenvectors respectively.}
 \label{CoarseApproxErrorSkyscrapers}
\end{figure}

With clearly structured problems, it can be visually seen that the GenEO coarse space
systematically picks up inclusions or channels in the parameter distribution. In Figure \ref{CoarseApproxErrorSkyscrapers} the coarse approximation error is shown
for a Darcy problem on a square domain. Dirichlet conditions are set to one at the top and zero at the bottom,
Neumann conditions are set at the remaining boundary and a high-contrast parameter distribution with jumps and channels
as shown on the left. We see that each inclusion has an effect
on the approximation error. Adding additional eigenvectors from each subdomain to the
coarse basis removes some of those error sources, the next eigenvectors pick up the skyscrapers and with only 4 eigenvectors per subdomain most channels are picked up.
A total of $16$ coarse basis functions is enough to almost entirely solve
the given problem.

\subsection{Demonstration of Robustness}

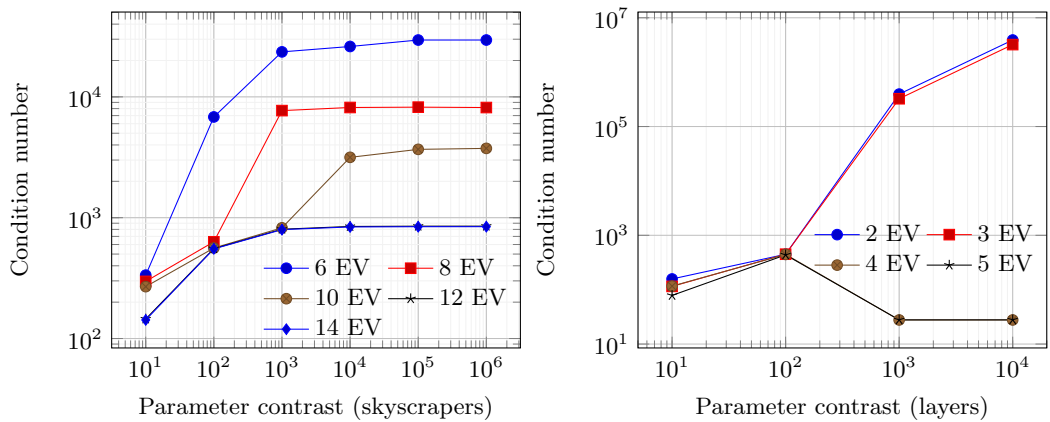
\begin{figure}[tbh]
\begin{tikzpicture}
\begin{loglogaxis}[xlabel=Parameter contrast (skyscrapers), ylabel=Condition number, grid=both,
    grid style={line width=.1pt, draw=gray!10},
    width=0.45\textwidth,
    major grid style={line width=.2pt,draw=gray!50},
    legend columns=2,
       legend style={draw=none,fill=none,at={(1,.3)}},
       legend cell align=left
    ]
\addplot table [x=Contrast, y=6 EV, col sep=comma] {geneo-skyscrapers-cond.csv};
\addlegendentry{6 EV};
\addplot table [x=Contrast, y=8 EV, col sep=comma] {geneo-skyscrapers-cond.csv};
\addlegendentry{8 EV};
\addplot table [x=Contrast, y=10 EV, col sep=comma] {geneo-skyscrapers-cond.csv};
\addlegendentry{10 EV};
\addplot table [x=Contrast, y=12 EV, col sep=comma] {geneo-skyscrapers-cond.csv};
\addlegendentry{12 EV};
\addplot table [x=Contrast, y=14 EV, col sep=comma] {geneo-skyscrapers-cond.csv};
\addlegendentry{14 EV};
\end{loglogaxis}
\end{tikzpicture}
\begin{tikzpicture}
\begin{loglogaxis}[xlabel=Parameter contrast (layers), ylabel=Condition number, grid=both,
    grid style={line width=.1pt, draw=gray!10},
    width=0.45\textwidth,
    major grid style={line width=.2pt,draw=gray!50},
    legend columns=2,
       legend style={draw=none,fill=none,at={(1,.4)}},
       legend cell align=left
    ]
\addplot table [x=Contrast, y=2 EV, col sep=comma] {geneo-layers-cond.csv};
\addlegendentry{2 EV};
\addplot table [x=Contrast, y=3 EV, col sep=comma] {geneo-layers-cond.csv};
\addlegendentry{3 EV};
\addplot table [x=Contrast, y=4 EV, col sep=comma] {geneo-layers-cond.csv};
\addlegendentry{4 EV};
\addplot table [x=Contrast, y=5 EV, col sep=comma] {geneo-layers-cond.csv};
\addlegendentry{5 EV};
\end{loglogaxis}
\end{tikzpicture}

\caption{Robustness of GenEO preconditioner}
\label{fig:geneo_robustness}
\end{figure}

Robustness with respect to parameter contrast can be demonstrated solving the same Darcy problem as in section \ref{GenEOStructuredProblems}. We choose a subdomain decomposition into 8 by 8 squares, a two-cell overlap region diameter
and a total of 800 $Q_1$ elements in each direction. Figure \ref{fig:geneo_robustness} (left) shows the resulting condition number for increasing contrast when setting
up a GenEO basis with various numbers of eigenvectors per subdomain. Clearly, the asymptotic robustness guaranteed by the analysis is achieved in practice.

When running the same setup with a parameter distribution of 40 horizontal equally thick layers, it becomes clear from Figure \ref{fig:geneo_robustness} (right) that robustness is achieved exactly at four eigenvectors per subdomain. That stems from the fact that
four coarse basis functions (together with the contribution of the one-level Schwarz method) are sufficient to represent the five layers contained in that subdomain. Similar relations can be observed with other strongly structured parameter distributions as well.

\subsection{Comparison to other solvers}
In this section we compare the performance of various preconditioned CG solvers. For this test we consider a flat composite plate of size $[0,100\mbox{mm}]\times[0,20\mbox{mm}]$. The laminate is made up of $12$ composite layers stacked in a sequence of different angles, refered to as a stacking sequence. The composite layers are seperated by very thin layers of resin. There is a large jump in material strength between the composite and resin layer and due to the rotated layers the anisotropy cannot be grid aligned. We discretise  with quadratic, 20-node serendipity elements to avoid shear locking and use full Gaussian integration. 

In Table~\ref{tab:Example01b} we compare the convergence of several iterative solvers. We record the condition number, the dimension of the coarse space $\mbox{dim}(V_H)$ if applicable and the number of CG iterations required to achieve a residual reduction of $10^{-5}$.  As expected the iteration counts increase steadily with the number of subdomains when no coarse space is used. In contrast, the iterations and the condition number estimates remain constant for the \textsc{GenEO} preconditioner as predicted by Thm.~\ref{ConditionNumberBound}. 

\begin{table}[h!]
\centering
\begin{tabular}{ccccccc}
\toprule
 & \multicolumn{2}{c}{1-level}  & \multicolumn{3}{c}{GenEO} &  \multicolumn{1}{c}{\textsc{BoomerAMG}} \\ \cline{2-7} 
$N_\text{core}$~~~~   & iter.       & $\kappa(\tmmathbf{A})$    & iter.   & $\kappa(\tmmathbf{A})$  & dim($V_H$)      & iter.   \\ \hline
4    & 89~      & 79,735         & ~~~16 ~~~~~  & 10    & 78  ~~            & 258        \\
8    & 97~      & 84,023         & ~~~15 ~~~~~  & 9     & 126   ~~          & 258         \\
16   & 107~     & 98,579         & ~~~16 ~~~~~  & 10    & 182     ~~        & 257         \\
32   & 158~     & 226,871        & ~~~16 ~~~~~  & 9     & 526   ~~          & 263        \\ 
\bottomrule
\end{tabular}
\caption{\label{tab:Example01b}Demonstration of performance of different preconditioners for a fixed problem size  of 30,000 DOFs and increasing the number of subdomains: Number of CG iterations (it), coarse space dimension ($\dim(V_h)$), an estimate of the condition number $\kappa(\tmmathbf{A})$.}
\end{table}

To demonstrate the efficiency of this approach we also compared with two different implementations of AMG. The implementation included in \texttt{dune-istl} has not originally been designed for composite application, and in its current form it does not
seem to be robust, especially in parallel \cite{dunecomp}.  We applied this solver  in its block version and measured strength of connection 
between two blocks for the aggregation procedure in 
the Frobenius norm. As a smoother we applied two SSOR iterations. In the test setup used here the \texttt{dune-istl} AMG converges very slowly or not at all, thus we do not include it in Table \ref{tab:Example01b}. We include results for \texttt{BoomerAMG} \cite{boomerAMG}. Here we retained the defaults for most parameters (HMIS coarsening without aggressive refinement levels and a hybrid Gauss-Seidel smoother). As recommended for elasticity problems we used blocked aggregation with block size corresponding to the spatial dimension. A strong threshold of $0.75$ was chosen after testing values in the range from $0.4$ to $0.9$. Due to a lower setup cost the \texttt{boomerAMG} solver is faster in actual CPU time than the GenEO solver in this small test case. However, for more complex geometries, \texttt{boomerAMG} does not perform very well and in our tests it does not scale beyond about $100$ cores in composite applications \cite{dune-cpc}.

\subsection{Industrially-motivated Example: Wingbox}
 In this section we describe an industrially motivated example in which we asses the strength of an airplane wingbox with a small localised wrinkle defect in one corner.
 Wrinkle defects often form during the manufacturing process and lead to strong local stress concentrations, which may cause premature failure \cite{dunecomp,San18}. More details on this test setup can be found in \cite{dune-cpc}.
 
\begin{figure}[t]
\centering
\includegraphics[width=0.47\textwidth]{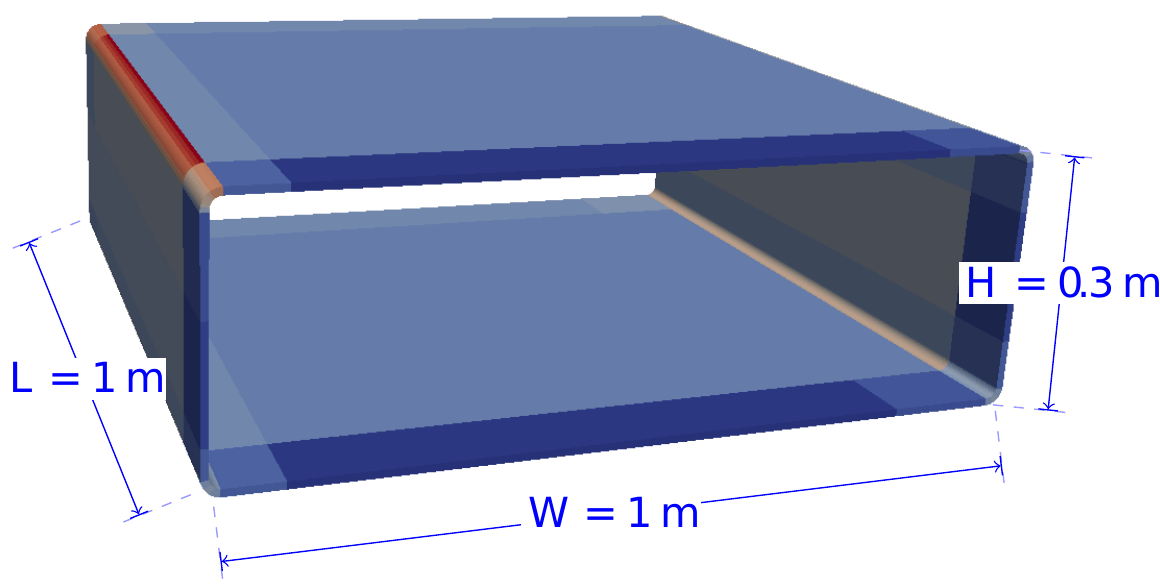}\hspace{1em}
\includegraphics[width=0.4\textwidth]{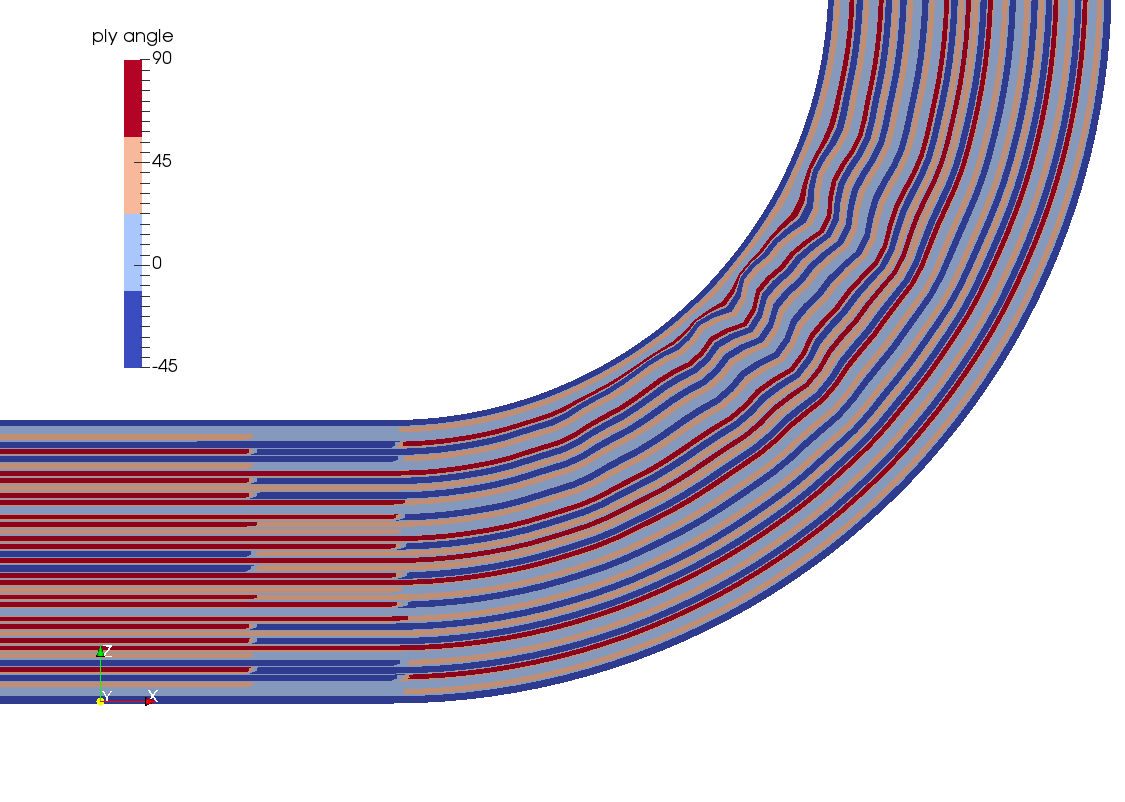}
\caption{Left: Geometry of the wingbox with dimensions; the colouring shows the number of eigenmodes used in \textsc{GenEO} in each of the subdomains of Setup 6 in Table~\ref{table-wingbox}. Right: The stacking sequence change around the corner containing a wrinkle. }
\label{fig:wingboxgeom}
\end{figure}

 We model a single bay of a wingbox of width $W = 1$m, height $H = 300$cm, length $L = 1$m and internal radius of the corners $15$mm, as shown in Figure \ref{fig:wingboxgeom} (left).  As in a typical aerospace application, the stacking sequence differs in the covers (top and bottom), corners and in the spar (sides). The changing stacking sequence is shown in Figure \ref{fig:wingboxgeom} (right). In total the laminate is made up of $39$ fibrous layers and $38$ resin layers. One of the corner radii contains a localised wrinkle with a parametrisation matching an observed defect in a CT-Scan of a real corner section, for further details see \cite{San18}.
 
Two forms of loading are applied, an internal pressure of $0.109$MPa, arising from the fuel, is applied to the internal surface and a thermal pre-stress induced by the manufacturing process is imposed. The influence of ribs that constrain the wingbox in the $y$ direction are approximated by clamping all degrees of freedom at one end and tying all other degrees of freedom at the other end using a multipoint constraint.

To demonstrate the effectiveness of the GenEO solver we carry out a weak scaling and a strong scaling experiment. The experiments in this section were performed using the UK national HPC cluster \textsc{Archer},  which consists of $4,920$ Cray XC30 nodes with two $2.7$ GHz, $12$-core E5-2697 v2 CPUs each. 

 For the weak scaling experiment we refine the mesh, doubling the number of elements as we double the number of cores.  Tab.~ \ref{table-wingbox} (left) contains the number of degrees of freedom, iteration numbers for the preconditioned CG, an estimate of the condition number of the discretisation matrix $\kappa(\tmmathbf{A})$, the dimension of the coarse space $\dim V_H$, as well as the total run time for each test. The weak scaling of the iterative CG solver with \textsc{GenEO} preconditioner is indeed almost optimal up to at least $15,360$ cores. 
 We note that as shown in Thm.~\ref{ConditionNumberBound} the condition number remains bounded. As expected we observe a clear connection between the number of iterations and the condition number.

Tab.~ \ref{table-wingbox} (right) shows a strong-scaling experiment. The iterative CG solver with GenEO preconditioner scales almost optimally to at least $11,320$ cores.  Memory constraints prevented tests with fewer than $2880$ cores. Tab.~\ref{table-wingbox} shows that the number of iterations indeed remains almost constant. The last column gives the parallel efficiency, it remains high up to $11,320$ cores.

\begin{table}
\begin{tabular}{cccccc}
\toprule
 $N_{core}$    & DOF & iter. & $\kappa(A)$ & $\dim(V_H)$ & Time (sec)\\
\midrule
 480  & $6.4\cdot10^6$ & 156 & 445 & 5025 & 734\\ 
 960 &  $1.3\cdot10^7$ & 154 & 421 & 7840 & 806\\ 
 1920 & $2.6\cdot10^7$& 152 & 322 & 18752 & 800\\ 
 3840 & $5.1\cdot10^7$& 144 & 287 & 29444 & 772\\ 
 7680 & $1.0\cdot10^8$& 132 & 303 & 50930 & 764\\ 
 15360 & $2.0\cdot10^8$& 102 & 245 & 94527 & 845 \\ 
\bottomrule
\end{tabular}
\begin{tabular}{cccccc}
\toprule
$N_{core}$  & elements/$N_{core}$ & $\mbox{dim}(V_H)$ & it. & Time (sec) & efficiency \\ \midrule
2880     & 3132              & 18843             & 167   & 2906       & 1.00\\
3840     & 2340              & 26333             & 153   & 1766       & 1.23\\
7680     & 2008              & 52622             & 132   & 1057       & 0.83\\
11320    & 1392              & 78233             & 162   & 706        & 1.01\\ 
\bottomrule
\\
\\
\end{tabular}
\caption{\label{table-wingbox}Parallel performance of the composites application on \textsc{Archer}. Left: Details of the weak scaling test. The number of elements per core was fixed at $2808$. Right: Strong scaling test demonstrating near optimal strong scaling up to at least $11,320$ cores.}
\end{table}

\section{Acknowledgements}
 This work was supported by an EPSRC Maths for Manufacturing grant (EP/K031368/1).
 This research made use of the Balena High Performance Computing
 Service at the University of Bath. This work used the ARCHER UK National Supercomputing Service (\url{http://www.archer.ac.uk}).

\bibliographystyle{splncs04}


\end{document}